# Pressure of a gas of underdamped active dumbbells

Marc Joyeux[(#)] and Eric Bertin[(+)]

*Université Grenoble Alpes, LIPHY, F-38000 Grenoble, France*
*CNRS, LIPHY, F-38000 Grenoble, France*

**Abstract:** The pressure exerted on a wall by a gas at equilibrium does not depend on the shape of the confining potential defining the walls. In contrast, it has been shown recently [A.P. Solon *et.al.*, Nat. Phys. **11**, 673 (2015)] that a gas of overdamped active particles exerts on a wall a force that depends on the confining potential, resulting in a net force on an asymmetric wall between two chambers at equal densities. Here, considering a model of underdamped self-propelled dumbbells in two dimensions, we study how the behavior of the pressure depends on the damping coefficient of the dumbbells, thus exploring inertial effects. We find in particular that the force exerted on a moving wall between two chambers at equal density continuously vanishes at low damping coefficient, and exhibits a complex dependence on the damping coefficient at low density, when collisions are scarce. We further show that this behavior of the pressure can to a significant extent be understood in terms of the trajectories of individual particles close to and in contact with the wall.

**Keywords:** Active matter, nonequilibrium pressure, underdamped dynamics, self-propelled dumbbells

**PACS**: 05.40.-a (Fluctuation phenomena, random processes, noise, and Brownian motion)

[(#)] email : marc.joyeux@univ-grenoble-alpes.fr
[(+)] email : eric.bertin@univ-grenoble-alpes.fr

**I - Introduction**

The pressure of fluids at thermodynamic equilibrium is a well-defined quantity, which can be estimated either from the mean force per unit area exerted by the constituent particles on confining walls or from the trace of the bulk stress tensor and satisfies an equation of state involving only bulk properties of the fluid, like temperature and density. Importantly, the existence of an equation of states implies that the value of the pressure does not depend on the microscopic detail of the interaction between the particles and the walls, be this interaction soft or hard, and torque-free or not. This is no longer the case for fluids far from equilibrium like active fluids, whose constituent particles are capable of autonomous dissipative motion like self-propulsion [1], and for which pressure [2-17], as well as stress [18] and other thermodynamic parameters like chemical potential [19], loose some of their standard thermodynamic properties. It has indeed been shown recently [8,9,16] that the pressure of active fluids is generally *not* a state function and that the average force exerted on confining walls by the fluid *does* depend on the detail of the interaction between the walls and the particles. It was furthermore shown that even for peculiar active fluids, like Active Brownian Spheres, which in two dimensions obey an equation of state when confined between torque-free walls [7], the introduction of a torque exerted by the wall is sufficient to prevent the existence of an equation of state since the pressure then depends on the wall potential [8].

Most of the results reported above were obtained in the limit of (explicitly [3, 4, 6-16] or effectively [2]) overdamped dynamics, that is, for self-propelled particles travelling essentially at fixed given speed, up to some positional noise. It turns out that the case of inertial particles characterized by a finite mass and a self-propulsion force and moving in a medium with finite but relatively small damping coefficient $\gamma$ has received little attention, although such underdamped systems may have a behavior closer to that of equilibrium systems and could thus be of conceptual interest for the development of thermodynamic theory of active systems. The purpose of the present work is to propose an underdamped model describing how the unusual properties of pressure reported in [8] emerge progressively from the characteristics of individual trajectories upon increase of the damping coefficient $\gamma$. In particular, it will be shown that, for low enough particle density, the evolution with increasing $\gamma$ is not monotonous and that the difference between the pressure exerted by the active particles on a wall with a large repulsion coefficient and the pressure exerted by the same particles on a wall with a weaker repulsion coefficient may change sign several times upon increase of $\gamma$. Analysis of the dynamics of the model further suggests that this pressure difference is governed for small values of $\gamma$ by trajectories with a large spin momentum, which hit the wall several times over short time intervals, and for larger values of $\gamma$ by the penetration depth of the active particle inside the mobile wall and the related duration of the interaction between the particle and the wall. Finally, it will be show that the influence of trajectories with a large spin momentum is progressively destroyed upon increase of the particle density, while the effect of the duration of the interaction between the particle and the wall is a more robust mechanism. This ultimately leads, for large enough particle density, to a monotonous onset of the effects reported in [8] upon increase of $\gamma$.

The remainder of this paper is organized as follows. The model is described in Section II and the results of simulations performed therewith are presented in Section III. These results are next interpreted in Section IV thanks to the analysis of the dynamics of single active particles. We finally discuss these results and conclude in Section V.

**II – Description of the model**

The model is schematized in Fig. 1. It consists of $N$ identical self-propelled dumbbells [5,20-26] moving in a 2-dimensional space and enclosed between fixed walls, which confine their motion inside an area with gross size $2L_x \times 2L_y$. A mobile wall of thickness $2e$ separates this area into two non-communicating chambers. The mobile wall can move along the $x$ axis while remaining parallel to the $y$ axis, the position of its median line being characterized by its abscissa $x_w$. Corners between any two walls, whether fixed or mobile, have the shape of a quarter of a circle of radius $r$, in order to avoid the accumulation of particles that occurs in square corners [6,27]. An equal number $N/2$ of dumbbells are enclosed in each chamber, each dumbbell $j$ being composed of two particles with respective positions $\mathbf{R}_{2j-1}$ and $\mathbf{R}_{2j}$ ($j=1,2,…N$) connected by a harmonic spring and separated at equilibrium by a distance $a$. The main feature of the model is that each particle experiences an active force directed from the tail of the dumbbell (the particle at position $\mathbf{R}_{2j-1}$) towards its head (the particle at position $\mathbf{R}_{2j}$). Besides this active force, each particle also interacts with the fixed and mobile walls through interaction potentials that vanish outside the wall and increase quadratically inside the wall, thus confining the particles inside each chamber. Finally, two neighboring particles that do not belong to the same dumbbell repel each other through a softcore potential, which vanishes for separations larger than $2a$ and increases quadratically for smaller ones. Note that no orientational noise was introduced in the model, in contrast with the Active Brownian Spheres and Run-and-Tumble Particles models [28-35], in order to avoid having to deal with (and discuss the results as a function of) the additional time scale related with the reorientation velocity. Since the particles interact with each other, they nonetheless experience collision-induced reorientations, which become very frequent at large dumbbell density.

More explicitly, the potential energy $V$ of the system (not including the active force) is written as the sum of three terms

$$V = V_s + V_{ev} + V_w , \qquad (\text{II-1})$$

where $V_s$ describes the internal (stretching) energy of the dumbbells, $V_{ev}$ the softcore repulsion between neighboring particles that do not belong to the same dumbbell, and $V_w$ the confining potential exerted by the walls on particles that tend to escape from the chambers. These three terms are expressed in the form

$$\begin{aligned}
V_s &= \frac{h}{2} \sum_{j=1}^{N} (\|\mathbf{R}_{2j-1} - \mathbf{R}_{2j}\| - a)^2 \\
V_{ev} &= \frac{h}{2} \sum_{k=1}^{2N-2} \sum_{\substack{m = \begin{cases} k+1 \, (k \text{ even}) \\ k+2 \, (k \text{ odd}) \end{cases}}}^{2N} H(2a - \|\mathbf{R}_k - \mathbf{R}_m\|) \times (2a - \|\mathbf{R}_k - \mathbf{R}_m\|)^2 \\
V_w &= \frac{h_L}{2} \sum_{k \in S_L} \|\mathbf{R}_k - \mathbf{p}(\mathbf{R}_k)\|^2 + \frac{h_R}{2} \sum_{k \in S_R} \|\mathbf{R}_k - \mathbf{p}(\mathbf{R}_k)\|^2 + \frac{h}{2} \sum_{k \in S_F} \|\mathbf{R}_k - \mathbf{p}(\mathbf{R}_k)\|^2 ,
\end{aligned} \qquad (\text{II-2})$$

where $H(r)$ is the Heaviside step function, which insures that particles that do not belong to the same dumbbell do not repel each other as long as their separation remains larger than $2a$. In the expression of $V_w$, $\mathbf{p}(\mathbf{R}_k)$ denotes the orthogonal projection of the vector coordinate $\mathbf{R}_k$ of a particle that has penetrated inside a wall on the surface of this wall (see Fig. 1), so that $\|\mathbf{R}_k - \mathbf{p}(\mathbf{R}_k)\|$ represents the penetration depth of this particle inside the wall. $S_L$, $S_R$, and $S_F$ furthermore denote the sets of particles that at the considered time $t$ have penetrated inside the mobile wall coming from its left (L) and right (R) sides, and the set of particles that have penetrated inside fixed (F) walls, respectively. Note that, for the sake of simplicity, the dumbbell harmonic spring, softcore repulsive potential, and fixed wall repulsive potential share the same force constant $h$.

The kinetic energy $T$ of the system is

$$T = \frac{m_w}{2}(\frac{dx_w}{dt})^2 + \frac{m}{2}\sum_{k=1}^{2N}\left\|\frac{d\mathbf{R}_k}{dt}\right\|^2, \quad \text{(II-3)}$$

where $m$ denotes the mass of each particle and $m_w$ the mass of the mobile wall.

Finally, the equations of motion of the system are written in the form

$$m\frac{d^2\mathbf{R}_k}{dt^2} = \mathbf{F}_k + m\gamma(v_0\mathbf{n}_{j(k)} - \frac{d\mathbf{R}_k}{dt})$$
$$m_w\frac{d^2 x_w}{dt^2} = F_w - m_w\gamma\frac{dx_w}{dt}, \quad \text{(II-4)}$$

($k=1,2,\ldots,2N$), where $\mathbf{F}_k$ is the force felt by particle $k$ resulting from the potential function $V$, $\gamma$ is the damping coefficient of the medium, $\mathbf{n}_j = (\mathbf{R}_{2j} - \mathbf{R}_{2j-1})/\|\mathbf{R}_{2j} - \mathbf{R}_{2j-1}\|$ the unit vector pointing from the tail to the head of dumbbell $j$, and $j(k)$ denotes the integer part of $(k+1)/2$. As anticipated above, Eq. (II-4) implies that, in addition to the force resulting from the potential function $V$, each particle $k$ is subject at any time to an intrinsic force $m\gamma v_0 \mathbf{n}_{j(k)}$, which is oriented from the tail to the head of the dumbbell it belongs to. Note that, for isolated particles (*i.e.* $\mathbf{F}_k = 0$ at any time), the stationary solution of Eq. (II-4) consists of rectilinear trajectories travelled at constant velocity $v_0$. Moreover, in the absence of collisions, the characteristic time for the alignment of the velocity vector of a particle along the tail-to-head axis of the dumbbell is $1/\gamma$.

The main motivation for considering dumbbells instead of simple point-like or spherical particles is that dumbbells naturally contain an axis, which is useful to define the self-propulsion force, and they also naturally give rise to a torque exerted by the walls, without adding any extra interactions on top of the potentials. These useful properties somehow reduce the number of arbitrary functions or parameters in the model.

Two points may be worth emphasizing. First, Eq. (II-4) does not conserve momentum, nor does it include any explicit coupling to a momentum-conserving medium, as is also the case for the Active Brownian Spheres and Run-and-Tumble Particles models [28-35]. Consequently, it is best suited to describe particles moving on a surface that acts as a momentum sink, like crawling cells [36] or colloidal rollers [37] and sliders [38]. Note however that such systems often have a large damping coefficient, while we allow in our model the damping coefficient to be small. Moreover, Eq. (II-4) implies that the medium surrounding the dumbbells contributes to the damping of the motion of the mobile wall and of the dumbbells but does not directly contribute to pressure forces (its contribution to pressure is only indirect, through its action on dumbbell dynamics). The wall is therefore assumed to be permeable to this medium and the pressure exerted by active dumbbells must be considered as an osmotic pressure [5,7]. Note also that some of the numerical simulations reported in [5] were done with inertial dumbbells, but comparisons with the present model are not straightforward because the model of [5] used a Nosé-Hoover thermostat, instead of a simple viscous friction term to dissipate the energy injected by the self-propulsion force.

For the purpose of numerical integration, the derivatives in Eq. (II-4) were discretized according to standard Verlet-type formulae and the equations of evolution subsequently recast into the form

$$\mathbf{R}_k^{(n+1)} = \frac{4}{2+\gamma \Delta t} \mathbf{R}_k^{(n)} - \frac{2-\gamma \Delta t}{2+\gamma \Delta t} \mathbf{R}_k^{(n-1)} + \frac{2 (\Delta t)^2}{m (2+\gamma \Delta t)} \mathbf{F}_k^{(n)} + \frac{2 v_0 (\Delta t)^2}{2+\gamma \Delta t} \gamma \mathbf{n}_{j(k)}^{(n)}$$

$$x_w^{(n+1)} = \frac{4}{2+\gamma \Delta t} x_w^{(n)} - \frac{2-\gamma \Delta t}{2+\gamma \Delta t} x_w^{(n-1)} + \frac{2 (\Delta t)^2}{m_w (2+\gamma \Delta t)} F_w^{(n)},$$

(II-5)

where superscripts $(n-1)$, $(n)$, and $(n+1)$ indicate the time steps at which the quantity is evaluated, and $\Delta t$ is the integration time step.

Most simulations were performed with the following set of geometrical parameters: $a = 1$, $L_x = L_y = 100$, $e = 8$ and $r = 20$, but some simulations were performed with $L_x = L_y = 300$, in order to check the importance of finite size effects (see below). Masses were set to $m = 0.5$ and $m_w = 2$, the mass of the mobile wall being thus of the same order of magnitude as the mass of dumbbells. Force constants were given the values $h = h_L = 4$ and $h_R = 0.4$ to introduce a strong dissymmetry between the left and right sides of the mobile wall, while $v_0$ was set to 2 in all the simulations discussed below. Moreover, $\gamma$ was varied between 0.02 and 1 for most simulations, but some simulations were also performed with values of $\gamma$ as small as 0.0005 to investigate the Hamiltonian limit of the model at small particle number. Note that for $\gamma \approx 0.01$ the characteristic time for the alignment of the velocity vector of a particle along the tail-to-head axis of the dumbbell is of the same order of magnitude as the time it takes for the particle to cross the empty chamber at velocity $v_0$, while velocity alignment is about hundred times faster than crossing for $\gamma = 1$. Finally, most simulations were performed with $N = 50$, 500, or 5000 dumbbells. Assuming that each particle effectively consists of a disk of radius $a$, the surface coverage corresponding to each value of $N$ can be estimated from the following formula for the system with no internal excitation ($V = 0$):

$$\sigma = \frac{Na^2(\frac{4\pi}{3} + \frac{\sqrt{3}}{2})}{4L_y(L_x - e) - 2(4 - \pi)r^2} \ . \tag{II-6}$$

For $L_x = L_y = 100$, this formula leads to $\sigma = 0.7\%$, 7.0%, and 70.0% for $N = 50$, 500, and 5000, respectively. All simulations were performed with a time step $\Delta t = 0.001$, which was checked to be small enough even for $N = 5000$.

Movies showing the evolution of the system over time windows of 400 time units are provided in the Supplemental Material [39]. They help visualize the effect of increasing dumbbell density on the collision and diffusion rates (movies S1 to S3), as well as the effect of increasing the damping coefficient $\gamma$ on the general characteristics of individual trajectories (movies S4 to S6).

## III – Evolution of pressure with $\gamma$ and $N$

This section is devoted to the presentation of the main results obtained with the model described above. To start with, let us first illustrate with a figure the crucial fact that the pressure exerted by an active fluid on a surface does depend on the microscopic details of the interactions between constituent particles and the surface, while this is not the case for fluids at thermodynamic equilibrium. Equilibrium may be recovered by considering a fluid of Brownian particles. We therefore start by comparing the case of active dumbells with that of equilibrium, Brownian dumbbells. To this end, let us note that, while Eqs (II-4) and (II-5) describe an active fluid, it actually suffices to replace the active force by random noise to let the equations describe an usual Brownian fluid. More explicitly, the following equations

$$\begin{aligned}\mathbf{R}_k^{(n+1)} &= \frac{4}{2 + \gamma \Delta t} \mathbf{R}_k^{(n)} - \frac{2 - \gamma \Delta t}{2 + \gamma \Delta t} \mathbf{R}_k^{(n-1)} + \frac{2(\Delta t)^2}{m(2 + \gamma \Delta t)} \mathbf{F}_k^{(n)} + \frac{2 v_0 (\Delta t)^2}{2 + \gamma \Delta t} \sqrt{\frac{\gamma}{\Delta t}} \boldsymbol{\xi}_k^{(n)} \\ x_w^{(n+1)} &= \frac{4}{2 + \gamma \Delta t} x_w^{(n)} - \frac{2 - \gamma \Delta t}{2 + \gamma \Delta t} x_w^{(n-1)} + \frac{2(\Delta t)^2}{m_w(2 + \gamma \Delta t)} F_w^{(n)} , \end{aligned} \tag{III-1}$$

where the $\boldsymbol{\xi}_k^{(n)}$ are random vectors with components extracted from a Gaussian distribution with zero mean and unit variance, describe a set of Brownian particles with mean squared velocity $v_0^2$ placed in the confinement device shown in Fig. 1 and subject to the internal potential $V$. If an equal number of dumbbells are placed in each confinement chamber and the equations of motion are integrated according to (III-1), one consequently expects that the position of the mobile wall will oscillate around an average zero abscissa, corresponding to equal areas for both chambers, whatever the values of the force constants $h_L$ and $h_R$ that characterize the interactions of the dumbbells with the left and right sides of the mobile wall. It can be checked in Fig. 2 that this is indeed the case for a simulation performed with a total number $N = 50$ of dumbbells, damping coefficient $\gamma = 1$, and force constants in a 1:10 ratio between the two sides of the mobile wall ($h_R = 0.4$ against $h_L = 4$). Also plotted on the same figure is the time evolution of the mobile wall position for the same system, except that the 50 dumbbells are now assumed to be active ones obeying Eq. (II-5) instead of Eq. (III-1). It is seen that, in this latter case, the mobile wall oscillates around a position that is displaced by more than $L_x/2$ towards negative abscissae and that the average area of the right confinement chamber is consequently more than three times larger than that of the left chamber. This reflects the fact that, for a given dumbbell density, the dumbbells exert a larger average force on the right side of the mobile wall than on its left side, so that the wall moves towards the left till the decrease in density in the right chamber and the increase in density in the left chamber compensate for the more efficient particle-wall interactions on the right side. The observation of such displacement of the mobile wall separating the two chambers filled with an equal number of dumbbells will be the essential tool of this study aimed at understanding the effect of the number of dumbbells $N$ and the damping coefficient $\gamma$ on the properties of the active fluid.

Results shown in Fig. 2 were obtained for a damping coefficient $\gamma = 1$, that is, in the case where the alignment of the velocity vector of each particle along the axis of the dumbbell occurs on a time scale much shorter than the time it takes for the dumbbell to cross the chamber at velocity $v_0$ and also much shorter than the characteristic time interval between two successive collisions (provided that the dumbbell density is low enough). This regime is actually qualitatively similar to the overdamped regime investigated in previous studies [2-16] and one may wonder what happens beyond this regime, that is, when the characteristic alignment time increases and eventually becomes as large as the crossing time. The answer to this question is provided in Fig. 3, which shows the evolution with $\gamma$ of the average relative position of the mobile wall, $<x_w>/L_x$, for $N = 50$, 500, and 5000 dumbbells. This figure is actually the central result of this paper. Most striking are, of course, the oscillations that are clearly observed at low $\gamma$ for $N = 50$ dumbbells. These oscillations indicate that the detail of the interactions between the particles and the wall and the average force exerted on each side of the wall vary significantly over small variations of $\gamma$. Increasing the number of dumbbells appears to decrease the amplitude of these oscillations, but the largest peak towards positive values of the wall abscissa around $\gamma = 0.2$ is still observed for $N = 500$ dumbbells, resulting in a somewhat unexpected non-monotonous evolution of the mobile wall average position. Finally, these oscillations are totally damped for $N = 5000$ dumbbells and the wall simply moves progressively towards the left with increasing values of $\gamma$. Understanding and rationalizing these observations will be the purpose of the following Section of this paper. In the remainder of the present Section, we will establish a couple of additional important results.

First, the average force per length unit (*i.e.* the pressure) averaged over both sides of the mobile wall fluctuating around its average position is roughly proportional to the density of dumbbells $\rho$ and the damping coefficient $\gamma$, as may be checked in Fig. 4. Here $\rho$ is taken as the density averaged over the two chambers, that is

$$\rho = \frac{N}{4L_y(L_x - e) - 2(4 - \pi)r^2} \ . \tag{III-2}$$

It is indeed seen in Fig. 4 that the points for $N = 500$ and $N = 5000$ nearly superpose and that $<F>/(\rho L_y)$ increases almost linearly in the range $0 \leq \gamma \leq 1$ for these two values of $N$, while the plot for very low dumbbell density ($N = 50$) undulates more widely around the linear evolution as a function of $\gamma$.

In contrast, no such simple dependence on $N$ emerges from the plots of the evolution of the particles mean squared velocity shown in Fig. 5. In this figure, $<v^2>$ denotes the squared modulus of the velocity of the particles, $\|d\mathbf{R}_k/dt\|^2$, averaged over particles, time, and trajectories. Head-on collisions with walls and other particles reduce the instantaneous velocity of a particle, which subsequently increases again with a characteristic time constant $1/\gamma$. If $\gamma$ is too small or $N$ too large, then particles are not able to achieve the limit velocity $v_0$ before next collision occurs. This is the reason why $<v^2>/v_0^2$ is clearly an increasing function of $\gamma$ and a decreasing function of $N$. One can try to quantify this trade-off between acceleration and collision forces by rescaling the curves corresponding to different densities onto a single master curve. As shown in the inset of Fig. 5, an approximate rescaling of the data can be achieved by plotting them as a function of the damping coefficient $\gamma$ divided by the square root of $N$, or equivalently by the square root of the density, which is nothing but the typical distance between particles.

The fact that in our model particles have a low speed at low values of the damping coefficient $\gamma$ is actually not trivial. Considering only the translational motion of isolated particles, one finds from Eq.(II-4) that the particles speed should stabilize to the value $v_0$ characterizing self-propulsion. According to this simple reasoning, the particle speed should thus be independent of the damping coefficient. However, one has to take into account the rotational and vibrational degrees of freedom of the dumbbells, as rotation and vibration of particles is triggered by collisions with the walls and with other particles. Dumbbell rotation and vibration generates more energy dissipation, because the rotation and vibration speeds may be large even if the translational velocity is small, and because the damping force acts on each particle composing the dumbbell (thus not only on its center of mass). Moreover, rotation makes the energy injection by the self-propulsion force much less efficient, since the direction of the force is continuously rotating. Hence for small damping $\gamma$, rotational motion is persistent and the self-propulsion force roughly behaves as a random force, since its orientation changes very fast. The amplitude of the self-propulsion force is $\gamma v_0$. Considering it as a random noise, the variance of this noise would be proportional to $(\gamma v_0)^2$. From standard Langevin equations, it is known that temperature is proportional to the ratio of noise variance to damping coefficient, and thus to $\gamma$, in the present case. The linear dependence of $<v^2>$ in the small $\gamma$, regime observed in Fig.5 can be qualitatively understood from this simple argument. The fact that it additionally depends on density probably results from the density-dependence of collisions rates. More frequent collisions, occuring for larger densities, lead to a stronger energy tranfer to the rotational and vibrational degrees of freedom, which are purely dissipative (no external energy injection occurs on rotational and vibrational degrees of freedom). As a result, dissipation is enhanced at high density, yielding a lower average kinetic energy, in agreement with the data shown in Fig. 5.

To conclude this section, let us finally mention that the results presented in Figs. 3-5 depend little on the exact geometry of the system, and particularly its finite size, even for a surface coverage as low as 7% (corresponding to $N = 500$ in Figs. 3-5). This can be checked in Fig. 6, which shows the evolution of the average position of the mobile wall as a function of the damping coefficient $\gamma$ for the system with $N = 500$ and $L_x = L_y = 100$ (same plot as in Fig. 3) and for the system with $N = 4912$ and $L_x = L_y = 300$, which share almost equal values of the density of dumbbells. It is seen in this figure that the two plots are indeed very close. In contrast, a stronger dependence of the average position of the wall on the geometry of the system (in particular the size of the container) may eventually occur for very low density of dumbbells (as for example $N = 50$ in Figs. 3-5) when $\gamma$ is close to 0.5. For smaller values of $\gamma$, the oscillations of the average wall position as a function of $\gamma$ are quite robust, and can be related to properties of individual dumbbells, as discussed below.

**IV – Contribution of individual dumbbell trajectories to the pressure**

The most striking feature of Fig. 3 is the non-monotonous evolution of the average position of the mobile wall for increasing values of $\gamma$, which is particularly marked for very low dumbbell density, leading to several oscillations in the plot of $<x_w>/L_x$ for $N = 50$, and is still clearly noticeable at larger density, resulting for example in a large incursion towards positive values of $x_w$ at $\gamma \approx 0.2$ for $N = 500$. Since the non-monotonous behavior is best observed for a small number of dumbbells, one may expect its origin to be understood by analyzing the properties of individual trajectories, that is, of a single dumbbell moving in a single confinement chamber. To this end, we modified the system by keeping only one confinement chamber (say, the left chamber) and placing only one active dumbbell therein. We furthermore assumed that the interactions between the dumbbell and the four walls still obey Eq. (II-4) but that collisions of the dumbbell against the right wall cause the confinement chamber to move as a whole towards the right, while preserving its shape and dimensions. In contrast, the three other walls of the chamber experience no recoil force during collisions with the dumbbell. We studied the properties of the modified system for increasing values of the damping coefficient $\gamma$ and different values of the force constant governing the repulsive interaction between the dumbbell and the right wall (which we call $h_w$ for the modified system, in order to avoid any confusion with the full system with two confinement chambers).

Fig. 7 shows the evolution of $\Delta x_w$, the average displacement of the confinement chamber per time unit, with increasing values of the damping coefficient $\gamma$, for $h_w = 0.4$ (i.e. the value of $h_R$ for the complete system) and $h_w = 4.0$ (i.e. the value of $h_L$ for the complete system). $\Delta x_w$ is a globally decreasing function of the damping coefficient $\gamma$ for both values of $h_w$, but broad fluctuations and narrow peaks are clearly superposed to this overall decrease. One such broad fluctuation is observed for $h_w = 0.4$ between $\gamma = 0.10$ and $\gamma = 0.15$, while another fluctuation is observed for $h_w = 4.0$ between $\gamma = 0.15$ and $\gamma = 0.30$. Due to these fluctuations, $\Delta x_w$ is substantially larger for $h_w = 0.4$ than for $h_w = 4.0$ in the range $0.10 \leq \gamma \leq 0.15$, while it is substantially larger for $h_w = 4.0$ than for $h_w = 0.4$ in the range $0.15 \leq \gamma \leq 0.30$. This observation correlates perfectly with the previous observation that, for the complete system with $N = 50$ dumbbells, the average position of the mobile wall, $<x_w>$, is negative in the range $0.10 \leq \gamma \leq 0.15$ and positive in the range $0.15 \leq \gamma \leq 0.30$ (see Fig. 3). Such correlations between the properties of the modified system with a single dumbbell and the complete system with $N = 50$ dumbbells are less clear above $\gamma = 0.30$ for reasons that will be explained below.

The origin of the relative increase in $\Delta x_w$ in certain ranges of $\gamma$ may be grabbed by examining more carefully the narrow peaks that are also observed in Fig. 7. When launching simulations with the corresponding values of the damping coefficient $\gamma$, dumbbells get very rapidly trapped along pseudo-periodic trajectories, which act as attractors. Such a pseudo-periodic trajectory is shown in Fig. 8 and Movie S7 [39] for $h_w = 0.4$ and $\gamma = 0.153$. When colliding with the right wall, the dumbbell acquires a spin momentum, which lets it come back rapidly against the wall and hit it again and again, resulting in a dramatic increase in $\Delta x_w$ for this particular value of $\gamma$. This result suggests that the broad fluctuations that surround these narrow peaks in Fig. 7 and correspond to relative increases in $\Delta x_w$ may be due to trajectories that share some resemblance with the pseudo-periodic trajectories (in the sense that the dumbbell acquires a large spin momentum during the collision with the right wall, which lets it come back and hit the wall several times over short time intervals) but are not strictly pseudo-periodic. If this hypothesis is correct, then the frequency of collisions between the dumbbell and the right wall should be comparatively larger in the corresponding range of values of $\gamma$ than outside this range. It can be checked in Figs. 9 and 10 that this is indeed the case. In these two figures, $\Delta x_w$, the mean displacement of the confinement chamber per time unit, is decomposed into the collision frequency, $f$, and the mean displacement of the chamber per collision, $\Delta x_w / f$. More precisely, Figs. 9 and 10 show the evolution with $\gamma$ of $f / \sqrt{\langle v^2 \rangle}$ and $\Delta x_w \langle v^2 \rangle / f$, respectively, where $\langle v^2 \rangle$ is the average squared velocity of the dumbbell (see Fig. S1 [39] for the plot of $\langle v^2 \rangle$ as a function of $\gamma$ for both values of $h_w$). The collision frequency is expected to be a linearly increasing function of the average velocity of the particle and it is consequently quite natural to plot $f/\sqrt{\langle v^2 \rangle}$ as a function of $\gamma$ to unravel subtler details. On the other hand, we found somewhat empirically that $\Delta x_w \langle v^2 \rangle / f$ evolves as an exponential function of $\gamma$ for both values of $h_w$. While it may perhaps be possible to find a rationale for this observation, we did not investigate this point further. The important point is that, besides the trivial effect associated with the velocity of the dumbbell, the frequency of the collisions between the right wall and the active dumbbell is indeed significantly larger in the range $0.0 \le \gamma \le 0.2$ for $h_w = 0.4$ and in the range $0.0 \le \gamma \le 0.4$ for $h_w = 4.0$ (see Fig. 9), thus confirming the importance of spin-induced multiple successive collisions between the active dumbbell and the right wall. Collisions between the right wall and the dumbbell are on average more efficient in pushing the confinement chamber towards the right for $h_w = 0.4$ than for $h_w = 4.0$ for values of $\gamma$ up to about 0.4 (see Fig. 10). As a result, increases in wall collision frequency due to spin momentum is likely to contribute to the oscillations that are observed in the average position of the mobile wall at low values of $\gamma$ (see Fig. 3).

Several remarks are in order here. First, the presence of narrow peaks above $\gamma = 0.4$ in Figs. 7, 9, and 10 indicates that attractive pseudo-periodic trajectories still exist at larger values of $\gamma$. However, the larger the value of $\gamma$, the faster the alignment of the velocity vector of the dumbbell along its tail-to-head axis, and the faster the damping of the spin momentum. As a consequence, pseudo-periodic trajectories become more and more rectilinear, dumbbells cross the confinement chamber several times during one pseudo-period, and pseudo-periods become larger and larger. This can be checked in Fig. 11 and Movie S8 [39], which show such a pseudo-periodic trajectory for $h_w = 4.0$ and $\gamma = 0.575$. Larger pseudo-periods imply in turn that the increase in the wall collision frequency (with respect to a random trajectory) is smaller compared to lower values of $\gamma$, which is reflected in the globally decreasing height of the narrow peaks with increasing $\gamma$ in Fig. 9. Spin momentum is of course also rapidly damped for all other trajectories in the same range of values of $\gamma$, which accounts for the decrease in the amplitude of broad fluctuations surrounding narrow peaks with increasing $\gamma$ in Fig. 9.

A second noteworthy remark is that, not only narrow peaks, but also narrow dips are observed in Figs. 7, 9 and 10. These narrow dips correspond to attractive periodic orbits, which may involve collisions with all the walls except for the right one and therefore do not contribute to the displacement of the confinement chamber. Like the pseudo-periodic trajectories hitting the right wall, these periodic orbits display large spin momenta at low $\gamma$ and become more and more rectilinear with increasing $\gamma$ (see for example Figs. S2 and S3 [39]). It may, however, be noted that the depth of these dips is generally smaller than the height of the narrow peaks and that they are usually not surrounded by broader fluctuations (relative decreases) of $\Delta x_w$ or $f$ even at low $\gamma$. This probably indicates that such periodic orbits are less stable and/or less efficient in attracting neighboring trajectories than the pseudo-periodic trajectories discussed above, but more work would be needed to ascertain if indeed - and why - this is the case.

Last but not least, the third remark deals with the robustness of the results and conclusions obtained so far for the modified model with a single active dumbbell placed in a single confinement chamber and their transferability to the complete model with $N$ active dumbbells placed in two confinement chambers separated by a mobile wall. As seen clearly in Movies S1 to S3 [39], the mean free path decreases rapidly from $N = 50$ to $N = 500$ and $N = 5000$. Indeed, dumbbells may cross the confinement chamber without colliding with another one for $N = 50$, while they always experience several collisions during the crossing over for $N = 500$, and their motion looks totally erratic due to very many collisions for $N = 5000$. Since for $N = 50$ the mean free path is substantially larger than the characteristic length of the trajectories responsible for the increase in wall collision frequency, it comes as no surprise that the subtleties of the dynamics of the modified model with a single dumbbell transfer well to the dynamics of the complete model with two chambers and $N = 50$ dumbbells, as already stated above. For $N = 500$, the mean free path is instead approximately equal to - or even somewhat shorter than - the characteristic length of trajectories looping back towards the mobile wall after a first collision with it, so that it may be expected that collisions between dumbbells interfere with the spin momentum mechanism. One accordingly observes in Fig. 3 that the amplitude of the displacement towards positive values of $<x_w>$ around $\gamma = 0.2$ is divided by a factor of about 3 upon increase of $N$ from 50 to 500, while this displacement is replaced by a shift towards slightly *negative* values of $<x_w>$ for $N = 5000$. Moreover, the weaker oscillations observed below $\gamma = 0.2$ for $N = 50$ are no longer observed for $N = 500$ and $N = 5000$. As a result, for $N = 5000$ the evolution of the mean position of the wall with increasing $\gamma$ just consists of a progressive displacement towards negative values of $<x_w>$, such as is also observed for $N = 50$ and $N = 500$ for values of $\gamma$ larger than 0.3.

Since this motion towards negative values of $<x_w>$ is not totally damped with increasing $N$, in contrast with the fluctuations arising from the spin momentum, it is likely due to a different mechanism, which will be described in the remainder of this Section.

A first indication concerning this second mechanism is provided by the observation that the exponential decrease of the mean displacement of the confinement chamber per collision with the active dumbbell, $\Delta x_w \langle v^2 \rangle / f$, which appears to be the rule for $h_w = 4.0$ in the range $0 \leq \gamma \leq 1$ and for $h_w = 0.4$ in the range $0 \leq \gamma \leq 0.6$, is replaced by an exponential *increase* for $h_w = 0.4$ and $\gamma \geq 0.7$ (see Fig. 10). Simultaneously to this change in the sign of the slope of $\Delta x_w \langle v^2 \rangle / f$, the mean duration of collisions between the right wall and the active dumbbell, denoted $\tau$, increases sharply with $\gamma$, as can be checked in Fig. 12. These two observations can be understood by realizing that the repulsive force and the torque exerted by the mobile wall on the active dumbbell depend on $h_w$ but not on the damping coefficient, while the active force increases linearly with $\gamma$. As a consequence, the penetration depth of the active dumbbell inside the mobile wall increases with $\gamma$ and it takes more and more time for the mobile wall to let the incoming dumbbell rotate and expel it out from the wall, this phenomenon being all the more marked for lower values of the wall repulsive force constant $h_w$. Fig. 10 shows that the net result is a strong increase of the mobile wall displacement per collision for lower values of $h_w$ compared to larger ones.

It may furthermore be expected that this mechanism be quite robust against an increase in the number of active dumbbells. Indeed, with increasing particle density, particles accumulate into the soft side of the wall (with a low repulsive force constant $h_w = 0.4$) due to the repulsion exerted by other particles in the system (see Movie S3 [39]), thereby increasing the duration of the interaction time between active particles and the mobile wall even further, while they are not able to accumulate in the hard side of the wall, which is characterized by a repulsive force constant $h_w = 4.0$ equal to that of overlapping particles.

Finally, the steady increase of the duration $\tau$ of the collisions over the whole $0 \leq \gamma \leq 1$ range (see Fig. 12) indicates that, while the associated effect on pressure difference becomes particularly efficient above $\gamma = 0.7$, such trend already exists for smaller values of $\gamma$. For small values of $\gamma$, the increase in wall collision frequency induced by the spin momentum is however predominant and masks the effect of collision duration for low dumbbell density. As discussed above, the spin momentum mechanism is, however, much less robust than the collision duration effect against the increase in the number of dumbbells, so that for large dumbbell density the latter mechanism prevails even for low values of $\gamma$. This is the reason why $<x_w>/L_x$ is a simple monotonous function of $\gamma$ for $N = 5000$ (see Fig. 3).

**V – Discussion and conclusion**

In this paper, we have studied numerically the behavior of the pressure in a gas of underdamped self-propelled dumbbells, considering mostly such a gas enclosed in a two-dimensional container with two chambers separated by a mobile asymmetric wall. Working with dumbbells has the advantage that any wall naturally exerts a torque on the dumbbells, leading to the situation studied in [8], apart from the fact that we are no longer considering the large damping limit. We have found that the displacement of the asymmetric wall, resulting from the unequal pressures exerted on both sides, varies continuously with the damping coefficient of the dumbbells, and goes to zero in the limit of zero damping coefficient. More strikingly, we also observed that in the low density regime, when collisions are scarce, the displacement of the wall (and thus the net force exerted on the wall at its initial position, when densities were equal on both sides) exhibits oscillations as a function of the damping coefficient. We have traced the origin of this non-monotonous behavior back to trajectories, which acquire a large spin momentum when colliding with the wall. The main effect of these specific trajectories is to enhance the collision frequency with the wall, since particles tend to be confined close to the wall. Moreover, the average interaction time between the active particles and the soft side of the mobile wall increases steadily with increasing values of the damping coefficient $\gamma$, which leads to a stronger average pressure being exerted on this side of the mobile wall for sufficiently large $\gamma$, whatever the density of active particles.

It is important to comment at this stage on the low damping limit of the present model, especially in view of the fact that the mean-square velocity goes to zero when the damping coefficient $\gamma$ goes to zero. One might thus think that this low damping limit is a trivial limit in which nothing moves. Let us first emphasize that we are not working at zero damping, but with small finite values of the damping coefficient $\gamma$. Hence, particles do move, although at a slow pace. Averages of physical observables are accordingly computed over larger and larger time windows as $\gamma$ is decreased. Yet, one might think that the fact that the force exerted by particles on the mobile wall becomes small when $\gamma$ is small (see Fig.4) is the reason why the displacement of the wall becomes small in this limit. However, it should be outlined that the mobile wall moves freely (in the sense that it is not confined by any potential except that of the container) and that its damping coefficient is equal to that of the particles. Hence although particles exert small forces, these forces still generate significant displacements. This is seen on Fig.7 (in the case of a single particle in a moving chamber), where the displacement per unit time of the mobile wall is shown as a function of $\gamma$. One sees that small values of $\gamma$ lead to larger displacements of the wall per unit time (while the number of collisions per unit time becomes lower). The mean-square velocity of the particle in this modified geometry is shown in Fig. S1 [39], and it behaves in a similar way as in the original system with $N$ particles. The reason for the large displacement of the wall is that a collision even with a small force generates a large displacement due to low damping of wall motion. Hence the fact that the moving wall remains close to the central position $x_w = 0$ is not due to the small amplitude of the force (or small speeds), but to the fact that the momentum transferred to the wall during a particle-wall collision does not depend on the shape of the repulsive potential in this limit.

In addition, the fact that the speed of particles plays no important role can be confirmed in an independent way. In the overdamped limit, the displacement of the wall can indeed be computed explicitly from Eq.(5) of Ref. [8] in the case of elliptic particles confined by an asymmetric mobile wall with harmonic confining potentials (a situation very close to ours for large $\gamma$). The position of the wall is found to be independent of particles speed and thus remains the same in the limit where the particle speed is very small.

We also emphasize that the oscillations observed at low density for the position of the wall as a function of the damping coefficient $\gamma$ indicate that the low-$\gamma$, inertial dynamics is far from trivial. Such oscillations would require further theoretical explanation. Among possible reasons that may account for these oscillations, we have pointed out the potential role played by individual trajectories with large spin momentum that generate frequent recollisions of a given particle with the wall.

Let us finally comment on the physical origin of the pressure, especially when damping is not too small. An important observation is that the instantaneous force exerted on the wall by a dumbbell depends only on the position of its center of mass (as soon as both particles composing the dumbbell are within the soft wall), and not on its orientation, because the wall potentials are harmonic. So the pressure depends essentially on the penetration depth of dumbbells in the wall, and on the duration of the interaction with the wall. For the simple and usual case of non-interacting active brownian particles, the duration of the interaction is determined by the time needed for the orientation of the particle coming into the wall to turn back due to the sole effect of angular diffusion, in the absence of any torque exerted by the wall. Hence in this situation, one may think of the pressure as resulting from particles pointing perpendicular to the wall. In our model, at large damping and large enough densities, particles accumulate into the soft wall due to the repulsive interaction of other particles in the system. In a sense, particles are "pushed into the wall" (see movie S3 [39]). Yet, when into the wall, particles are subjected to a torque that quickly aligns them with the wall. Hence we end up (again at large $\gamma$) with an accumulation into the wall of particles mostly aligned with the wall, but that nevertheless exert a pressure on the wall due to the repulsive wall potential.

Among several possible extensions of the present study, future work could focus on the effect of including noise in the dumbbell dynamics. In this case, the equilibrium situation does not correspond to a vanishing value of the damping coefficient, but rather to a finite value given by the fluctuation dissipation relation, in the case of white noise (for colored noise, equilibrium requires the introduction of a memory kernel in the damping term [40]). Hence regimes where the damping is smaller than at equilibrium could also be studied in this framework. Additional research directions could include the study of the effect of a low damping on transport of [41] or trapping by [42] mobile wedges, or on the sorting effect of particles in ratchet geometry [43]. All these effects, which are hallmarks of active particles dynamics, are expected to be weakened in the presence of a low friction. Comparison to driven dimers, that have been studied in the context of granular matter [44, 45, 46], could also be of interest to clarify the relations between self-propelled particles and driven granular systems.

**FIGURE CAPTIONS**

**Figure 1** : (color online) **(a)** Schematic diagram of a dumbbell, showing the two particles located at positions $\mathbf{R}_{2j-1}$ (tail) and $\mathbf{R}_{2j}$ (head), the string connecting them, and the active force applied to each particle and directed from the tail to the head of the dumbbell. **(b)** Schematic diagram of the confinement chambers. Fixed walls are shown as black solid lines and the mobile wall as red dotted lines. $x_w$ denotes the abscissa of the median line of the mobile wall. Also shown is the position $\mathbf{R}_k$ of a particle that has penetrated inside a fixed wall and its projection $\mathbf{p}(\mathbf{R}_k)$ on the surface of the wall. The repelling force exerted by the wall on this particle is proportional to $\|\mathbf{R}_k - \mathbf{p}(\mathbf{R}_k)\|$. The force constant associated with the repulsion potential on the left side of the mobile wall ($h_L$) and the right side of the mobile wall ($h_R$) are different.

**Figure 2** : (color online) Time evolution of the relative position of the mobile wall for $N=50$ active dumbbells (Eq. II-5, lower green trace) and $N=50$ Brownian dumbbells (Eq. III-1, upper red trace). The ordinate represents the abscissa of the wall $x_w$ averaged over time windows of length $10^4$ and divided by $L_x = 100$. Each trace corresponds to a single simulation performed with force constants $h_L = 4$ and $h_R = 0.4$ and damping coefficient $\gamma = 1$.

**Figure 3** : (color online) Evolution, as a function of the damping coefficient $\gamma$, of the average relative position $<x_w>/L_x$ of the mobile wall for $N=50$ (red solid line), $N=500$ (blue circles), and $N=5000$ (green squares) active dumbbells obeying Eq. (II-5). $L_x = L_y = 100$ for all simulations. Each point was obtained by averaging $x_w/L_x$ over 8 simulations and sufficiently long time windows to warrant uncertainties smaller than 0.01. The insert provides a zoom on the curve for $N=50$ dumbbells at low values of $\gamma$.

**Figure 4** : (color online) Evolution of $<F>/(\rho L_y)$ as a function of the damping coefficient $\gamma$ for $N=50$ (red solid line), $N=500$ (blue circles), and $N=5000$ (green squares) active dumbbells obeying Eq. (II-5). $<F>$ is the average magnitude of the force exerted by the dumbbells on both sides of the mobile wall fluctuating around its average position. $<F>$ was estimated from the same trajectories and time windows as $<x_w>/L_x$ in Fig. 3.

**Figure 5** : (color online) Evolution of $<v^2>/v_0^2$ as a function of the damping coefficient $\gamma$ for $N=50$ (red solid line), $N=500$ (blue circles), and $N=5000$ (green squares) active dumbbells obeying Eq. (II-5). $<v^2>$ is the average squared velocity of dumbbells estimated from the same trajectories and time windows as $<x_w>/L_x$ in Fig. 3. The inset shows a rescaling of the curves in the main figure, where $<v^2>/v_0^2$ is plotted as a function of $\gamma/\sqrt{N}$ instead of $\gamma$.

**Figure 6** : (color online) Evolution, as a function of the damping coefficient $\gamma$, of the average relative position $<x_w>/L_x$ of the mobile wall for the system with $N=500$ active dumbbells and $L_x=L_y=100$ (blue circles, same plot as in Fig. 3) and for the system with $N=4192$ active dumbbells and $L_x=L_y=300$ (brown lozenges). Each point was obtained by averaging $x_w/L_x$ over 8 simulations and sufficiently long time windows to warrant uncertainties smaller than 0.01.

**Figure 7** : (color online) Evolution of $\Delta x_w$ as a function of the damping coefficient $\gamma$ for $h_w=0.4$ (solid red line) and $h_w=4.0$ (dashed blue line) for the modified model with a single confinement chamber and a single active dumbbell enclosed therein. $\Delta x_w$ is the average displacement towards the right of the confinement chamber per time unit. This plot was obtained by integrating the equations of motion for $2\times10^{12}$ time steps and increasing $\gamma$ by $5\times10^{-13}$ at each time step. $\Delta x_w$ was subsequently computed over intervals of $\gamma$ of width $10^{-3}$.

**Figure 8** : (color online) Representation of a pseudo-periodic trajectory of the active dumbbell for the modified model with a single confinement chamber and $h_w=0.4$ and $\gamma=0.153$. The confinement chamber moves towards the right by equal increments each time it is hit by the dumbbell. Represented in this figure is only its "final" position corresponding to the "final" position of the dumbbell. See Movie S7 [39] for a movie of the same trajectory.

**Figure 9** : (color online) Evolution of $f/\sqrt{\langle v^2 \rangle}$ as a function of the damping coefficient $\gamma$ for $h_w = 0.4$ (solid red line) and $h_w = 4.0$ (dashed blue line) for the modified model with a single confinement chamber and a single active dumbbell enclosed therein. $f$ is the number of times the dumbbell hits the right wall per time unit and $<v^2>$ the average squared velocity of the dumbbell. The horizontal green dot-dashed line is just a guideline for the eyes. See the caption of Fig. 7 for computational detail.

**Figure 10** : (color online) Evolution of $\Delta x_w \langle v^2 \rangle / f$ as a function of the damping coefficient $\gamma$ for $h_w = 0.4$ (solid red line) and $h_w = 4.0$ (dashed blue line) for the modified model with a single confinement chamber and a single active dumbbell enclosed therein. $\Delta x_w / f$ is the average displacement towards the right of the confinement chamber per collision with the dumbbell, and $<v^2>$ the average squared velocity of the dumbbell. The green and brown dot-dashed lines are guidelines for the eyes aimed at emphasizing the overall exponential decrease of $\Delta x_w \langle v^2 \rangle / f$ with $\gamma$. See the caption of Fig. 7 for computational detail.

**Figure 11** : (color online) Same as Fig. 8, but for $h_w = 4.0$ and $\gamma = 0.575$. See Movie S8 [39] for a movie of the same trajectory.

**Figure 12** : (color online) Evolution of $\tau$, the mean duration of collisions between the right wall and the active dumbbell, as a function of the damping coefficient $\gamma$ for $h_w = 0.4$ (solid red line) and $h_w = 4.0$ (dashed blue line) for the modified model with a single confinement chamber and a single active dumbbell enclosed therein. See the caption of Fig. 7 for computational detail.

**Figure 1**

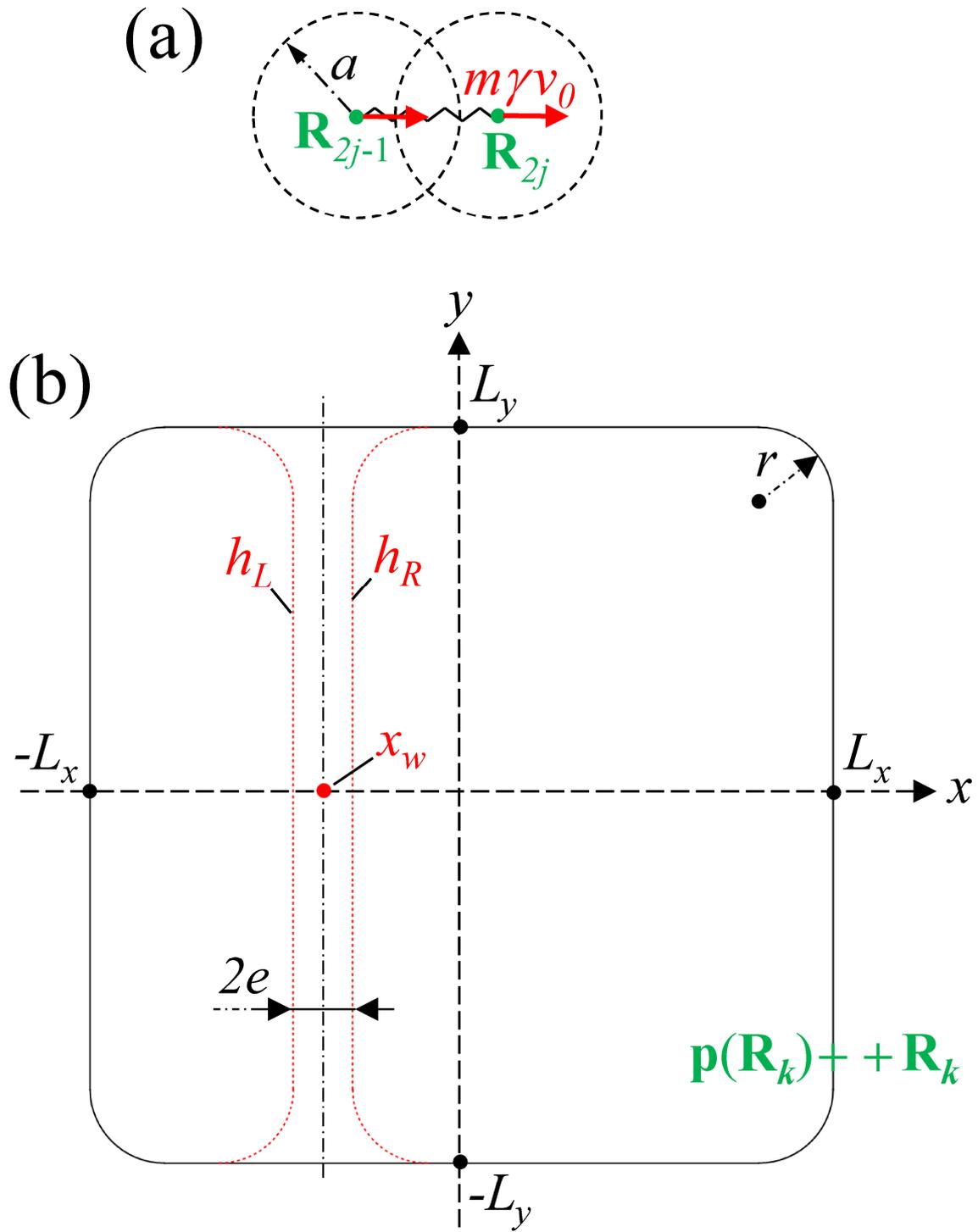

**Figure 2**

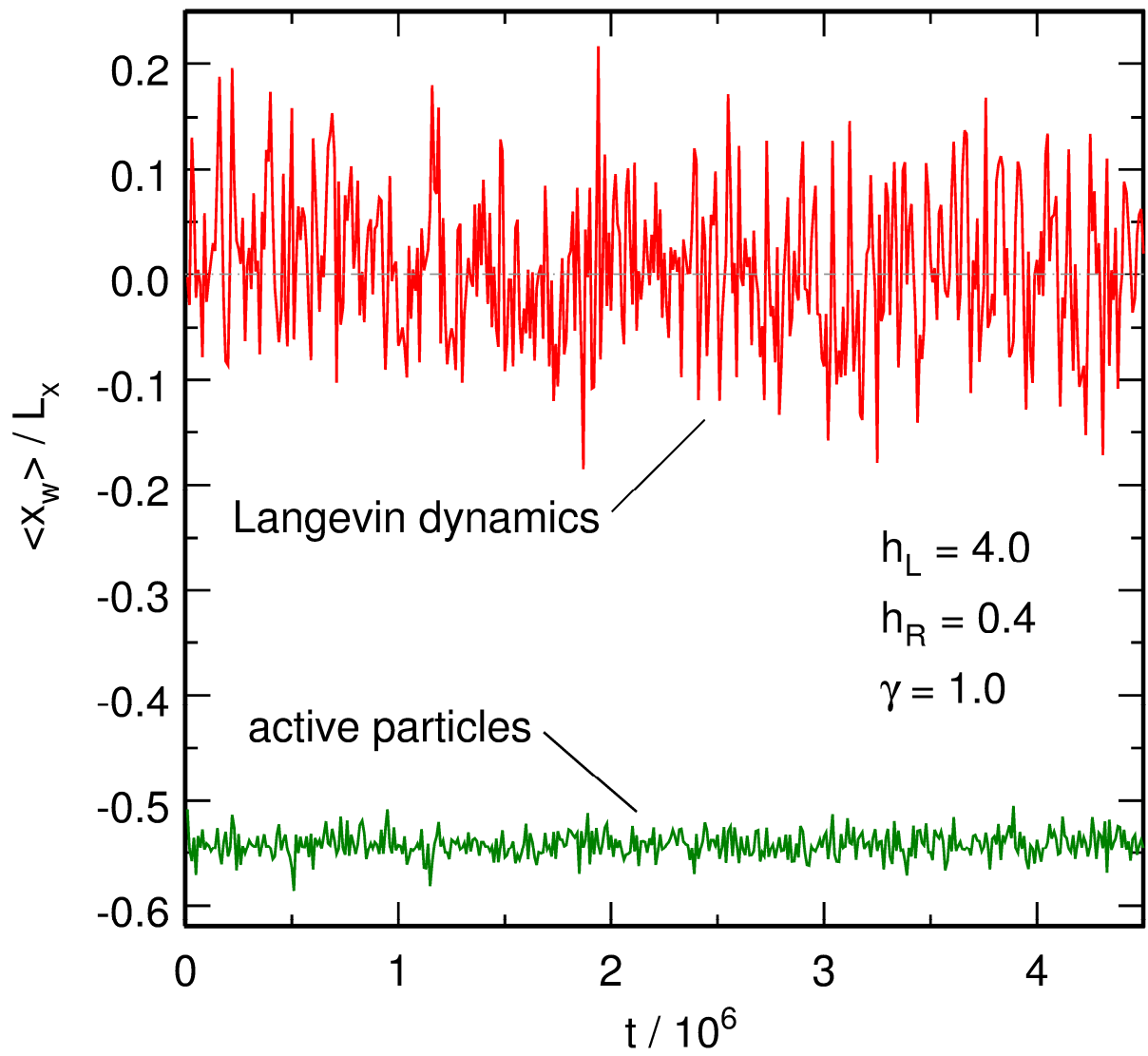

**Figure 3**

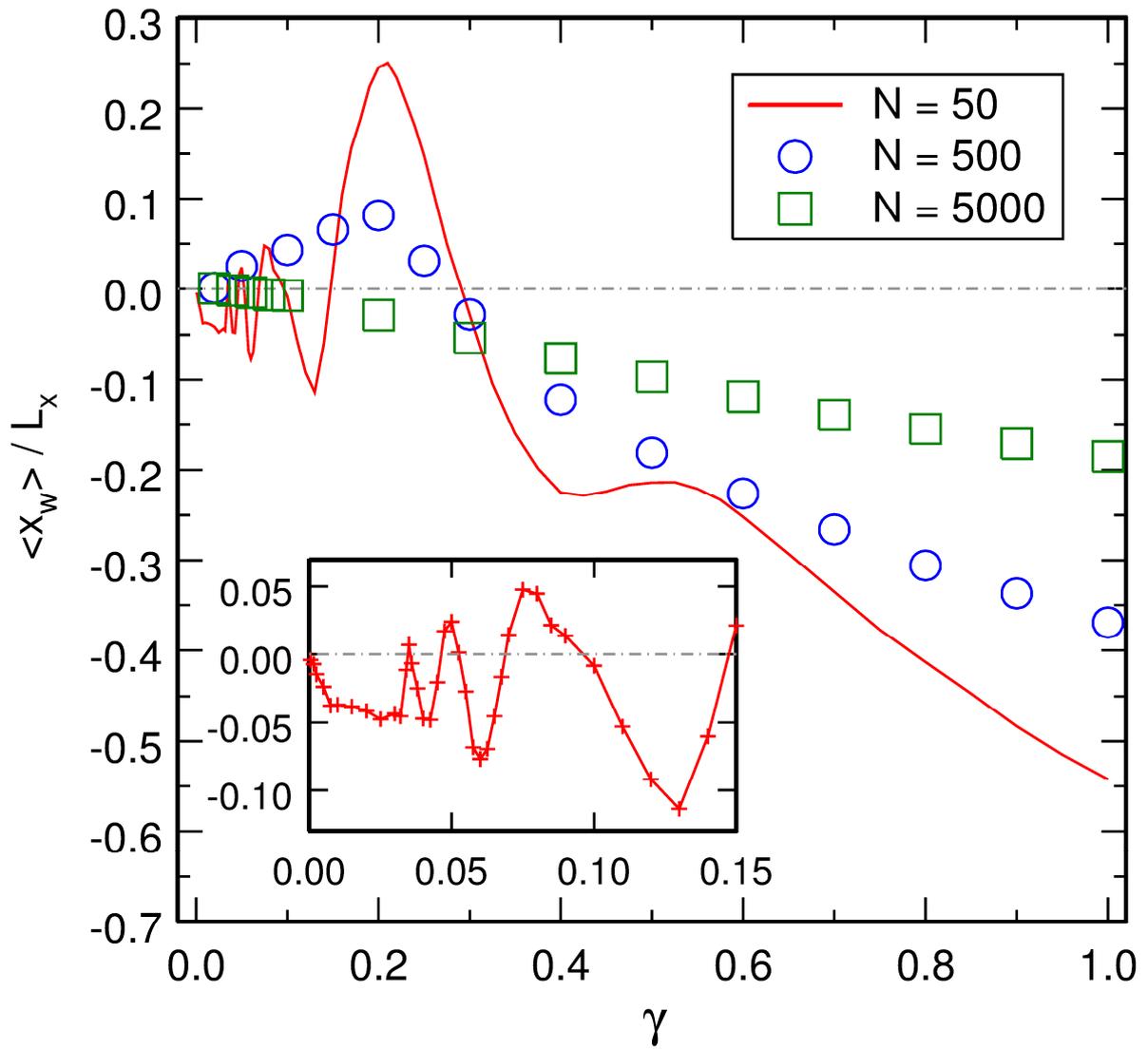

**Figure 4**

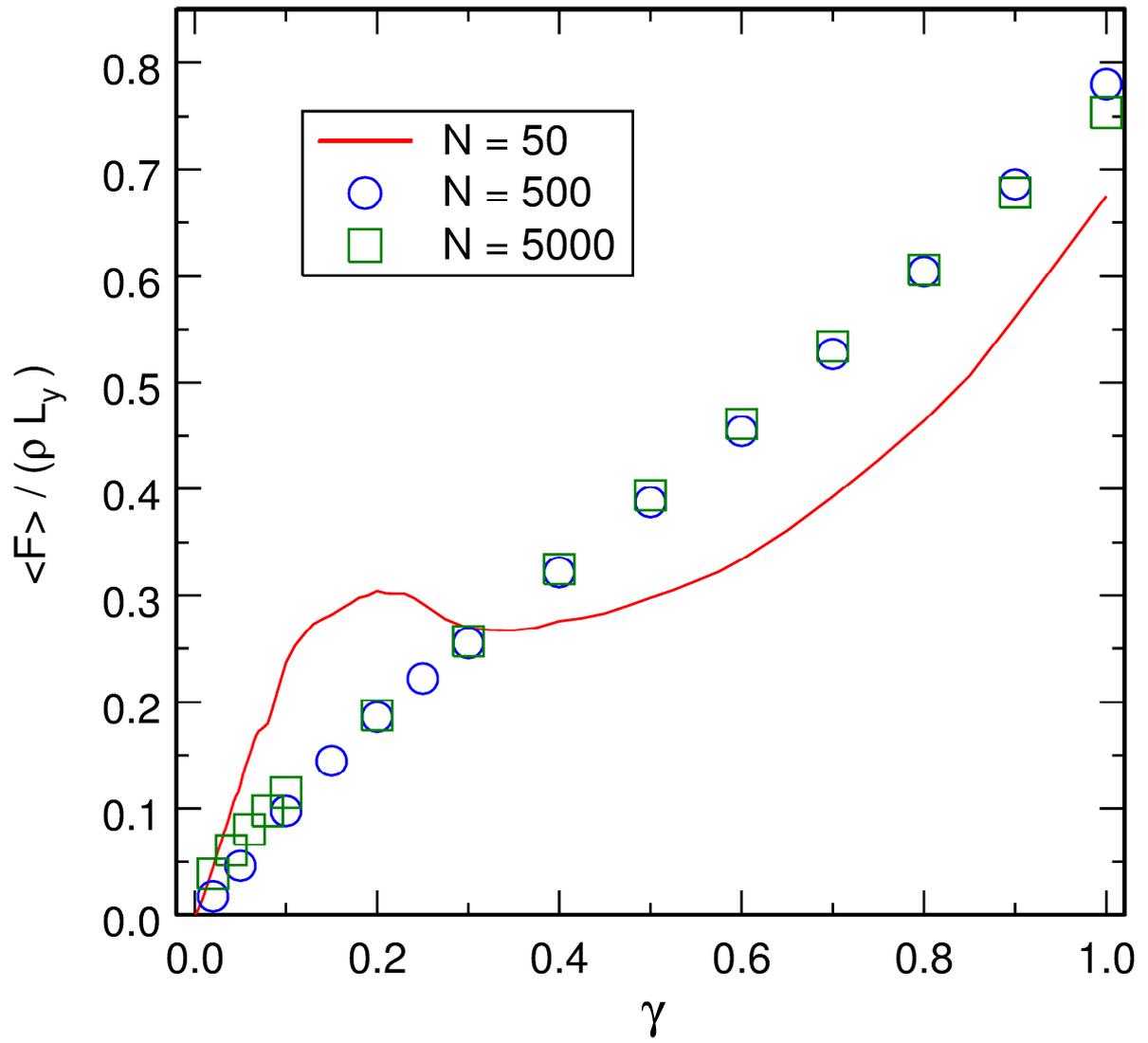

**Figure 5**

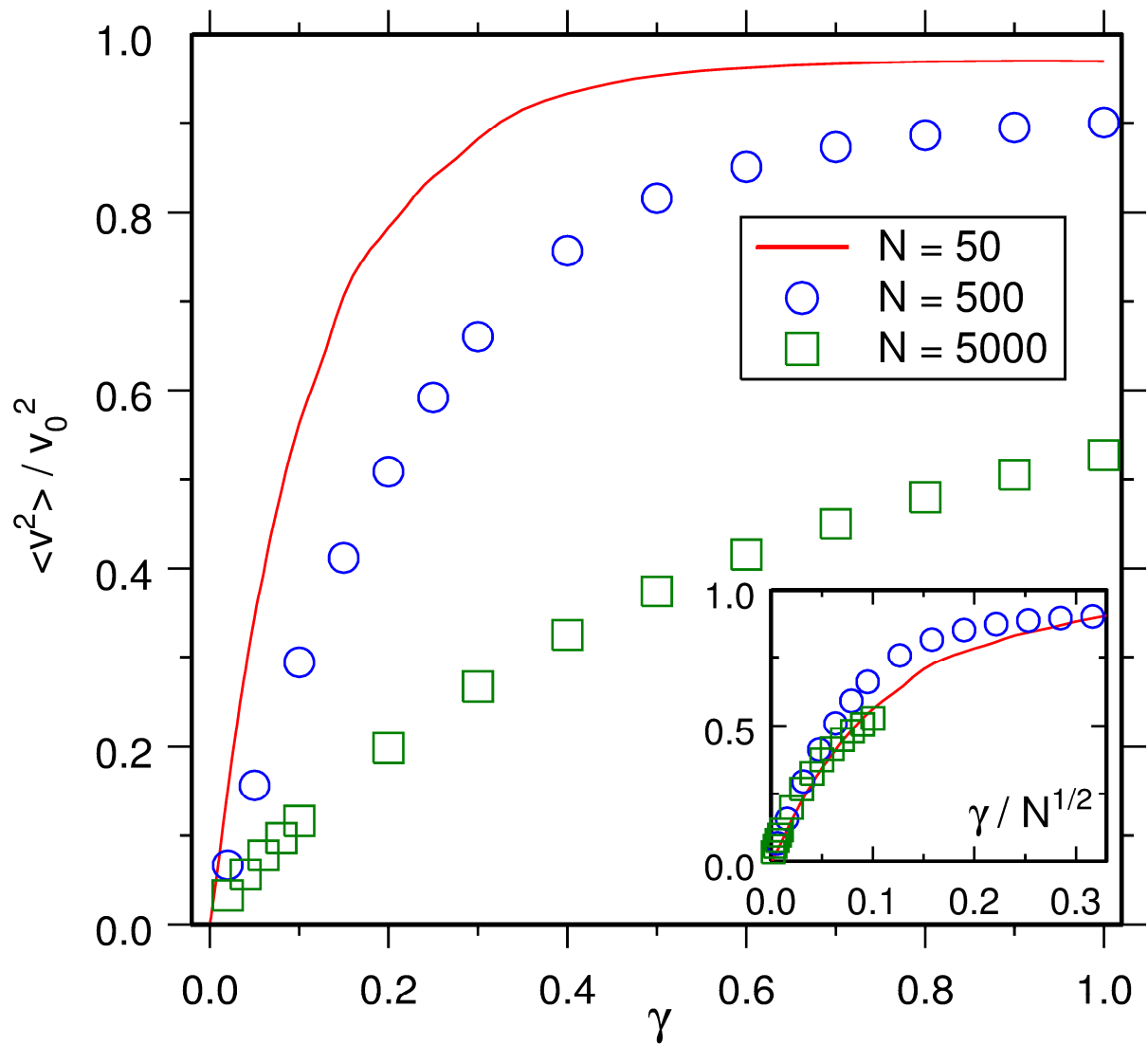

**Figure 6**

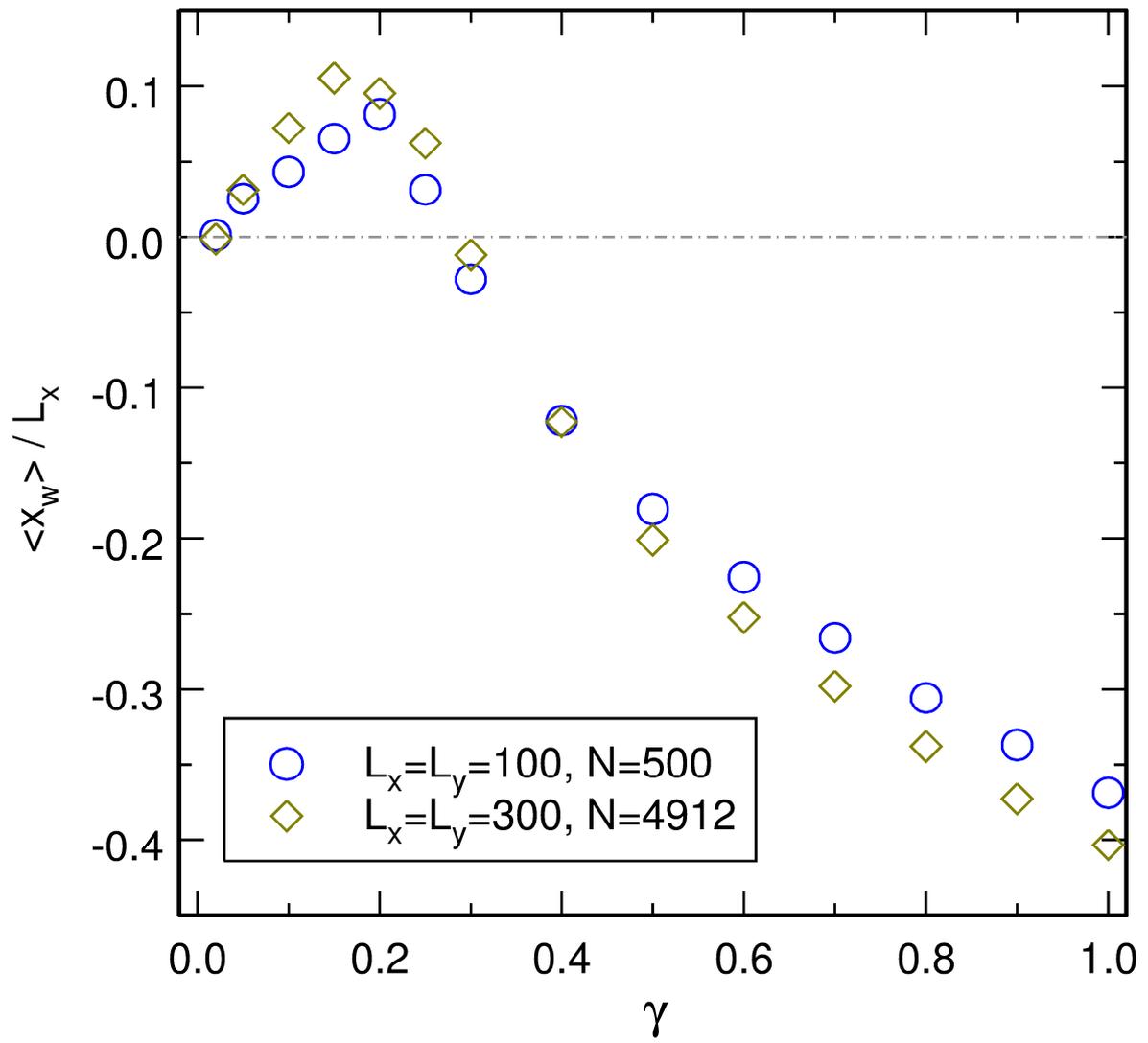

**Figure 7**

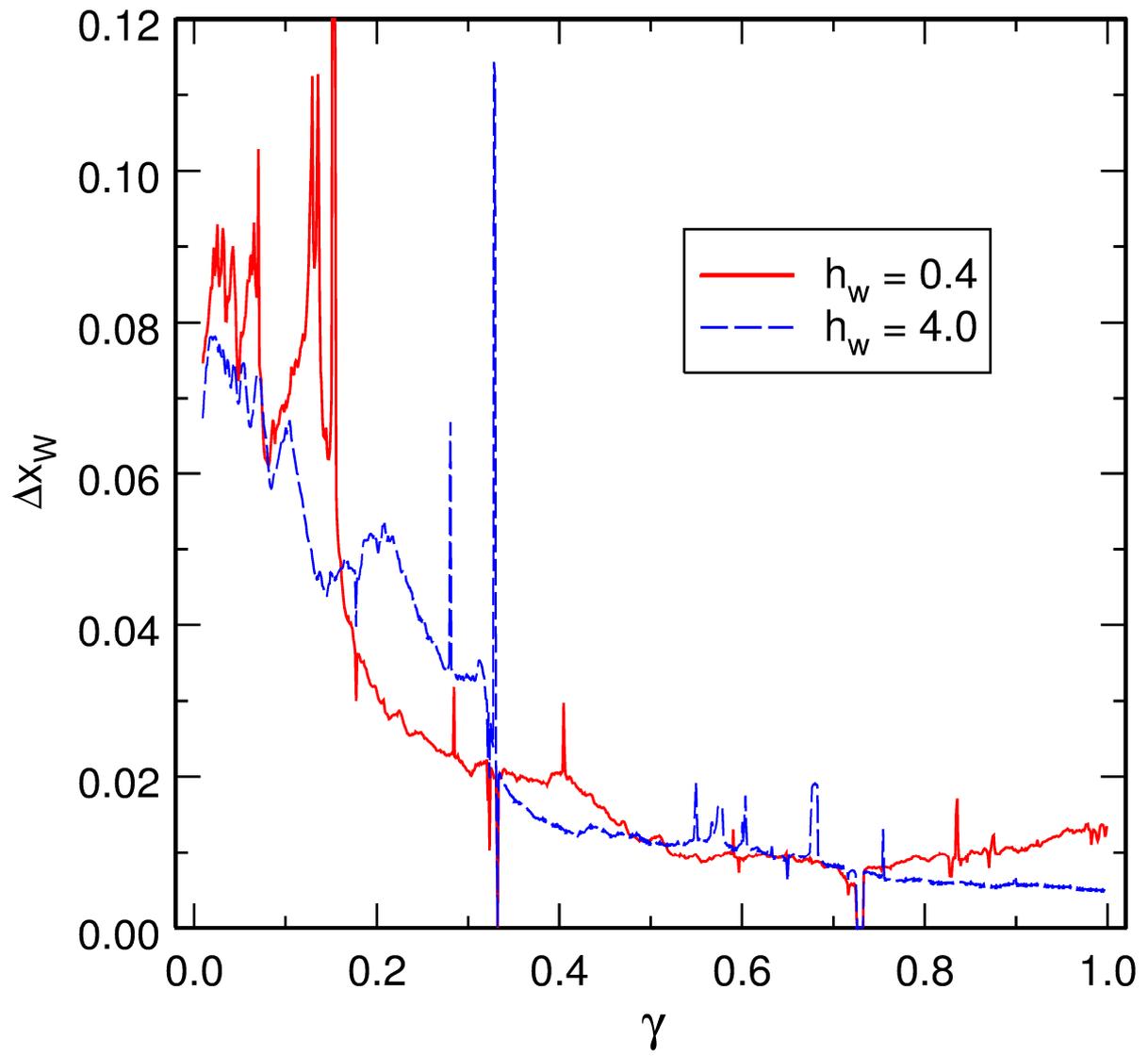

**Figure 8**

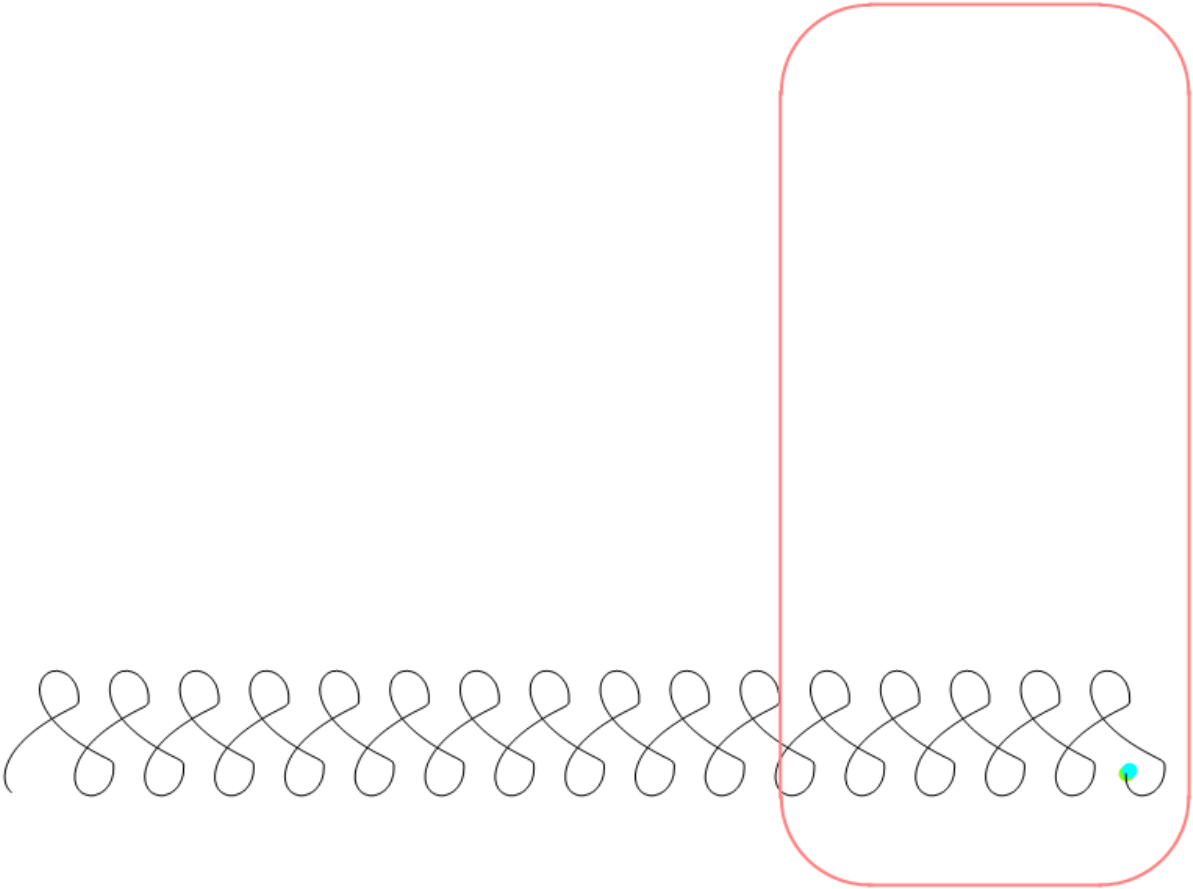

**Figure 9**

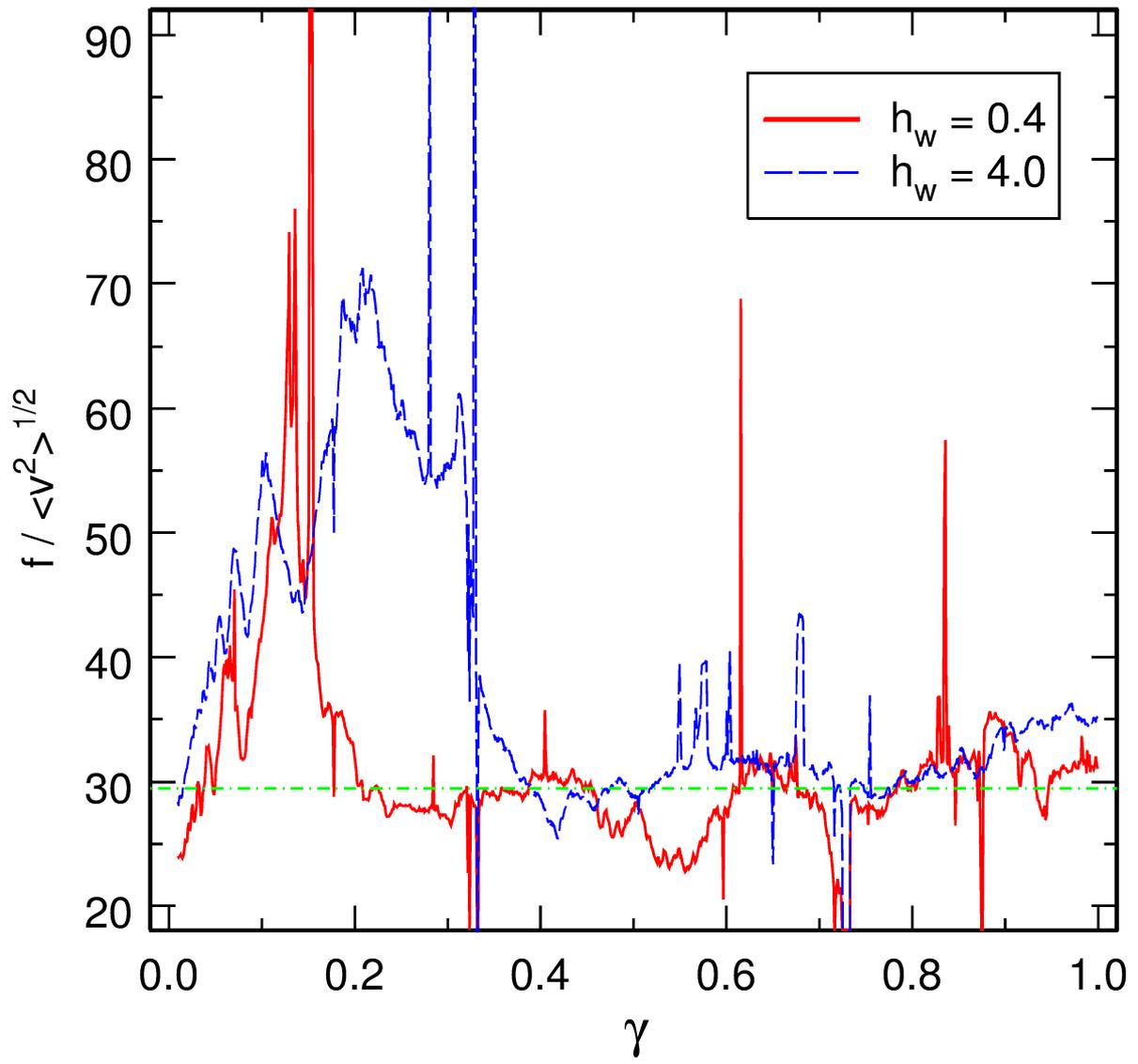

**Figure 10**

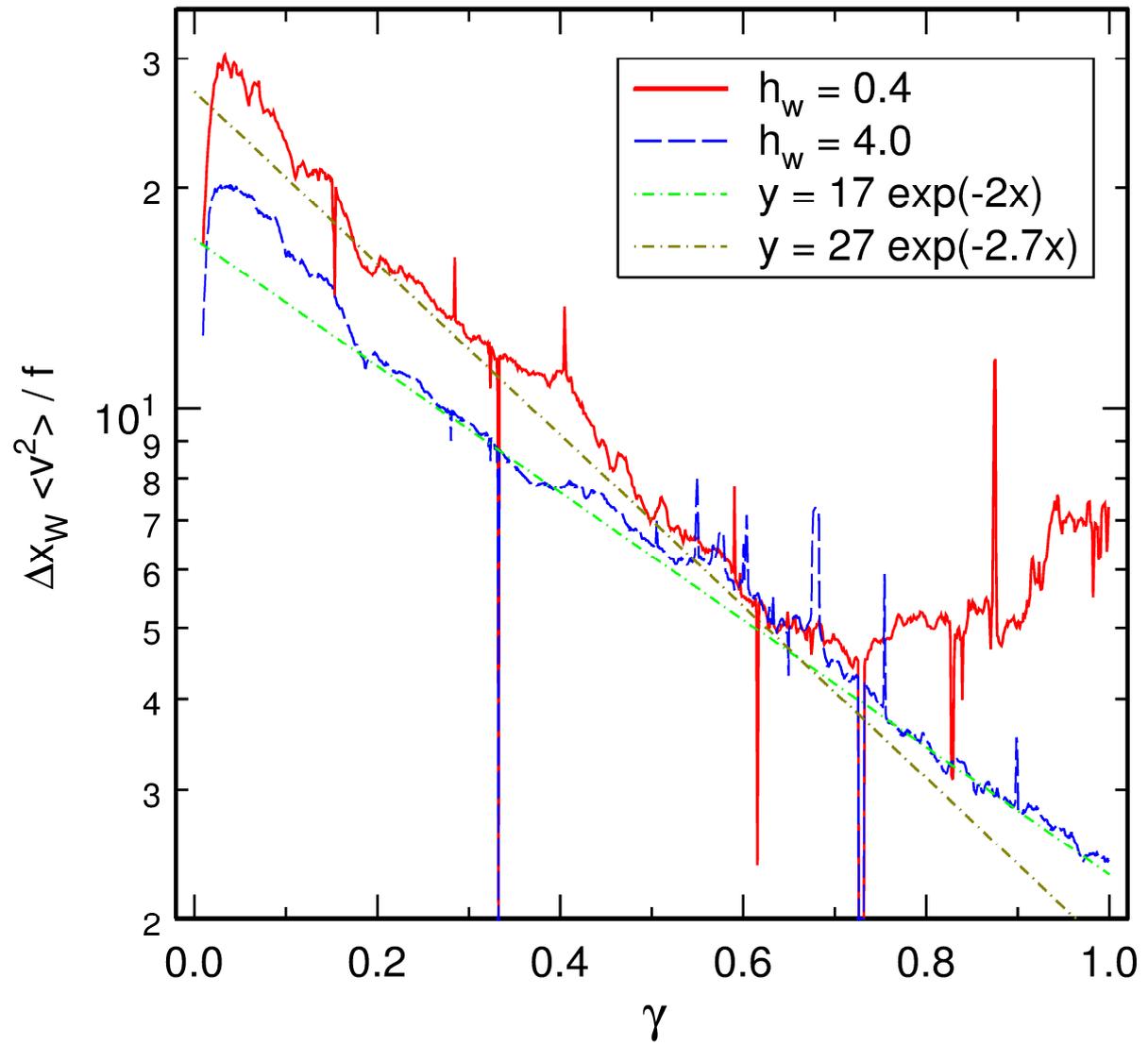

**Figure 11**

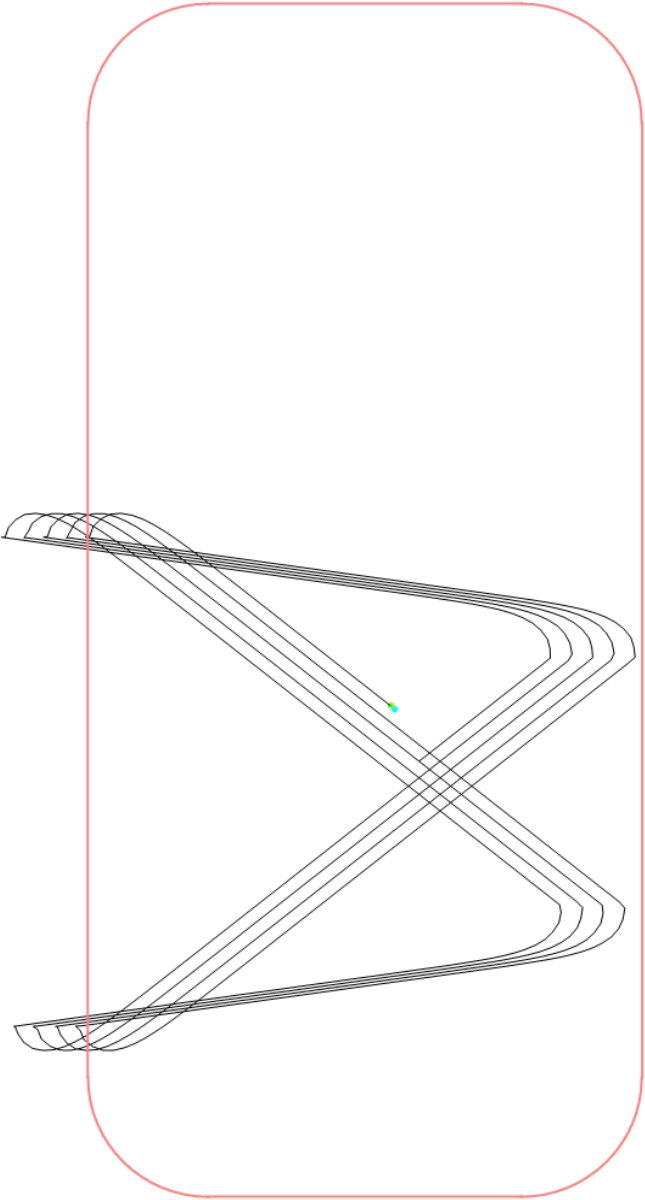

**Figure 12**

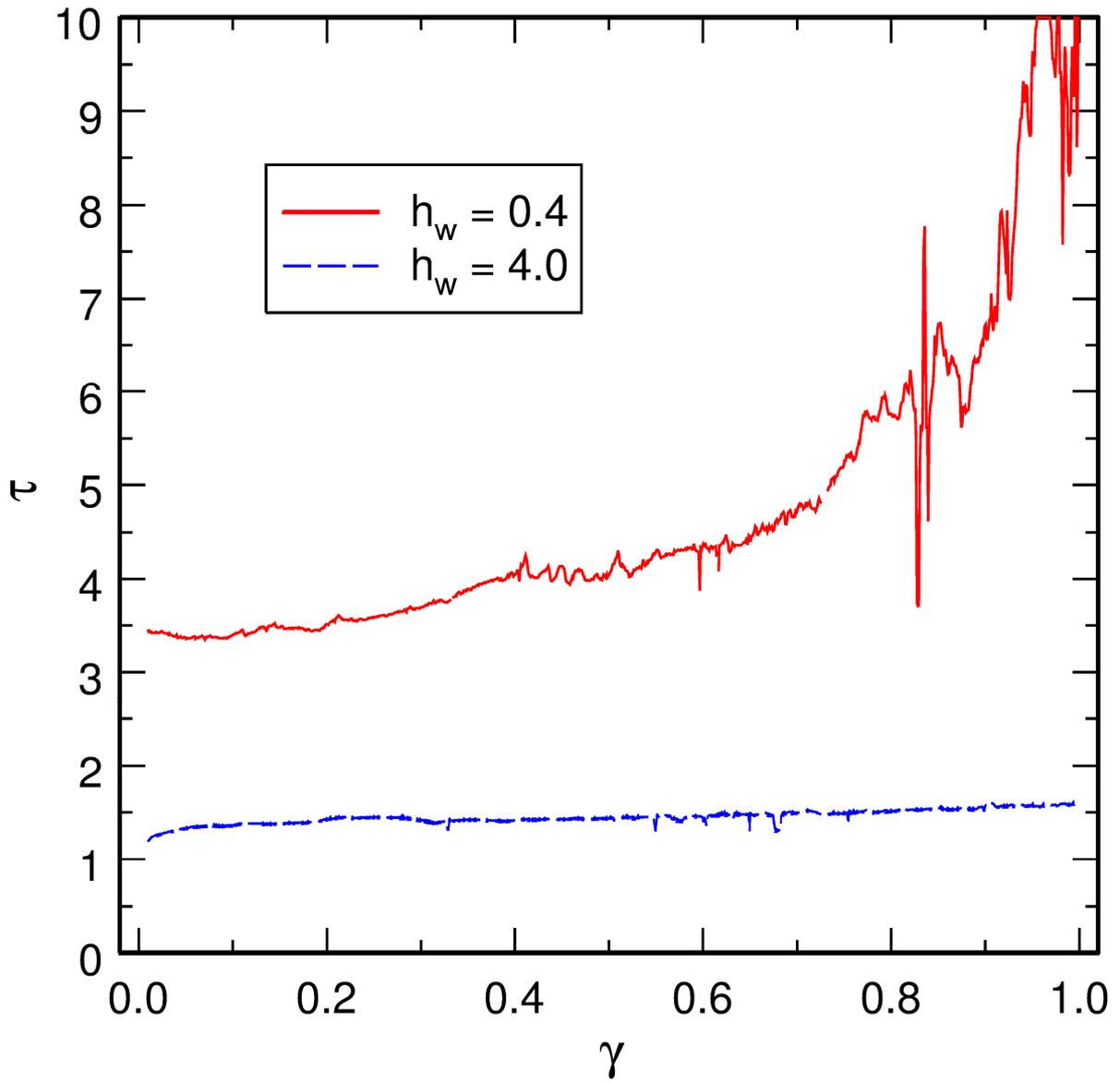

# CAPTIONS OF THE SUPPLEMENTAL MATERIAL

**Figure S1** : (.jpg file) Evolution of the relative mean squared velocity of the active dummbell as a function of the damping coefficient gamma for the modified model with a single confinement chamber and force constant h_w=0.4 (solid red line) or h_w=4.0 (dashed blue line). The dot-dashed green line is the plot of an exponential function with adjusted parameters, which was found empirically to reproduce satisfactorily the red and blue curves, and is used here merely as a guideline to the eyes. See the caption of Fig. 7 for computational details.

**Figure S2** : (.jpg file) Representation of a periodic orbit travelled by the active dumbbell for the modified model with a single confinement chamber, damping coefficient gamma=0.332, and force constant h_w=0.4.

**Figure S3** : (.jpg file) Same as Fig. S2, but for damping coefficient gamma=0.730 and force constant h_w=4.0.

**Movie S1** : (.avi file, 4 Mo) Movie showing the evolution of the system with N=50 active dumbbells, damping coefficient gamma=1, and Lx=Ly=100, over 400000 time steps, that is, a time window of length 400 time units. The movie is shown at 10 frames per second, two successive frames being separated by a time interval of 2000 time steps. Shown is the centre of mass of each dumbbell. The trajectory of one particle chosen randomly in each chamber is shown as a black line.

**Movie S2** : (.avi file, 8 Mo) Same as movie S1, but for N=500 active dumbbells.

**Movie S3** : (.avi file, 23 Mo) Same as movie S1, but for N=5000 active dumbbells.

**Movie S4** : (.avi file, 3 Mo) Same as movie S1, but for N=2 active dumbbells and damping coefficient gamma=0.01.

**Movie S5** : (.avi file, 3 Mo) Same as movie S4, but for damping coefficient gamma=0.1.

**Movie S6** : (.avi file, 3 Mo) Same as movie S4, but for damping coefficient gamma=1.0.

**Movie S7** : (.avi file, 2 Mo) Movie showing a pseudo-periodic trajectory of the dumbbell for the modified model with a single confinement chamber, damping coefficient gamma=0.153, and force constant h_w=0.4. This kind of trajectory (eventually translated along the *y* axis) acts as an attractor for all trajectories launched with the same values of the damping coefficient and force constant. The movie is shown at 10 frames per second, two successive frames being separated by a time interval of 2000 time steps.

**Movie S8** : (.avi file, 2 Mo) Same as Movie S7, but for damping coefficient gamma=0.575 and force constant h_w=4.0.

**Figure S1**

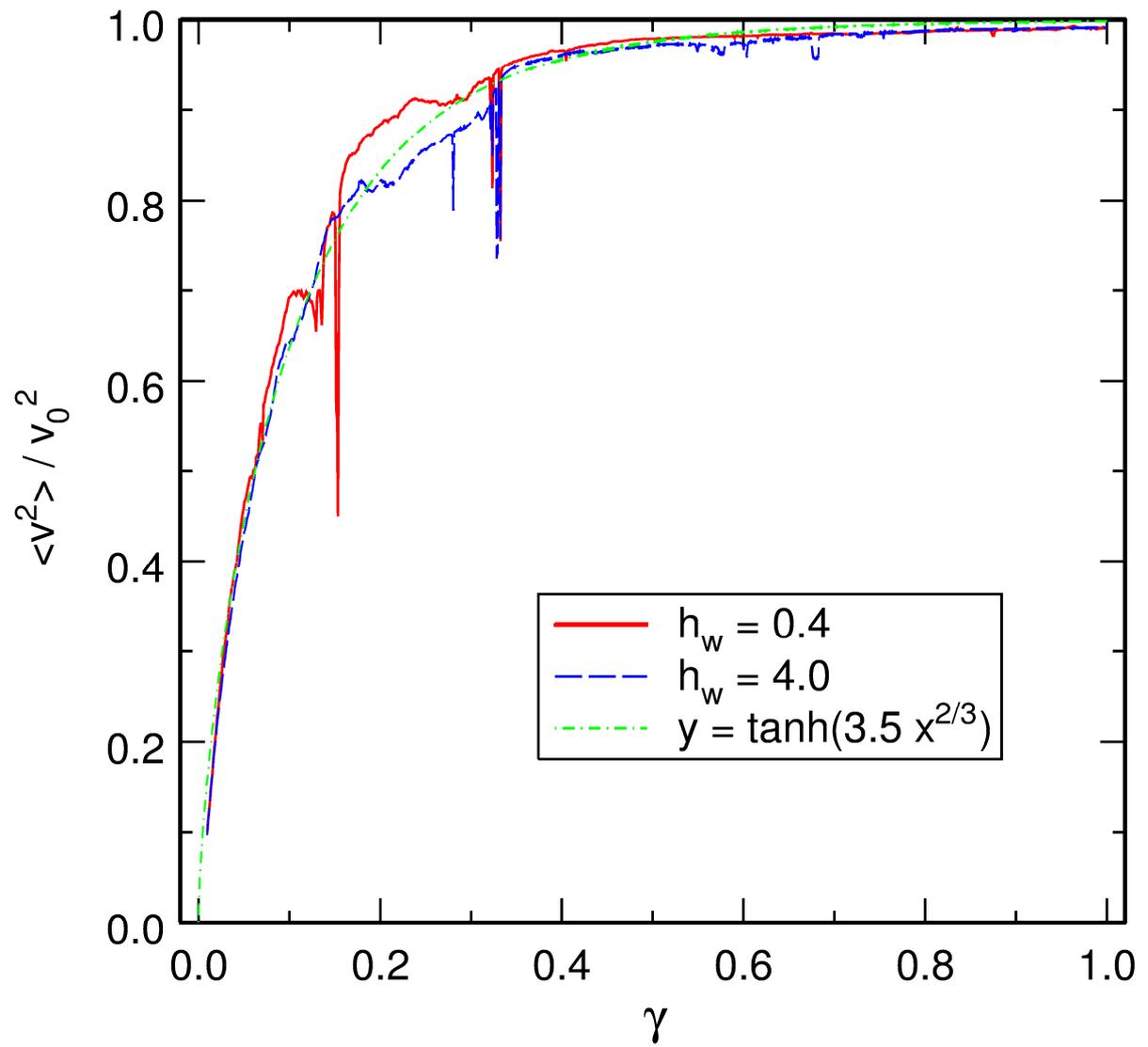

**Figure S2**

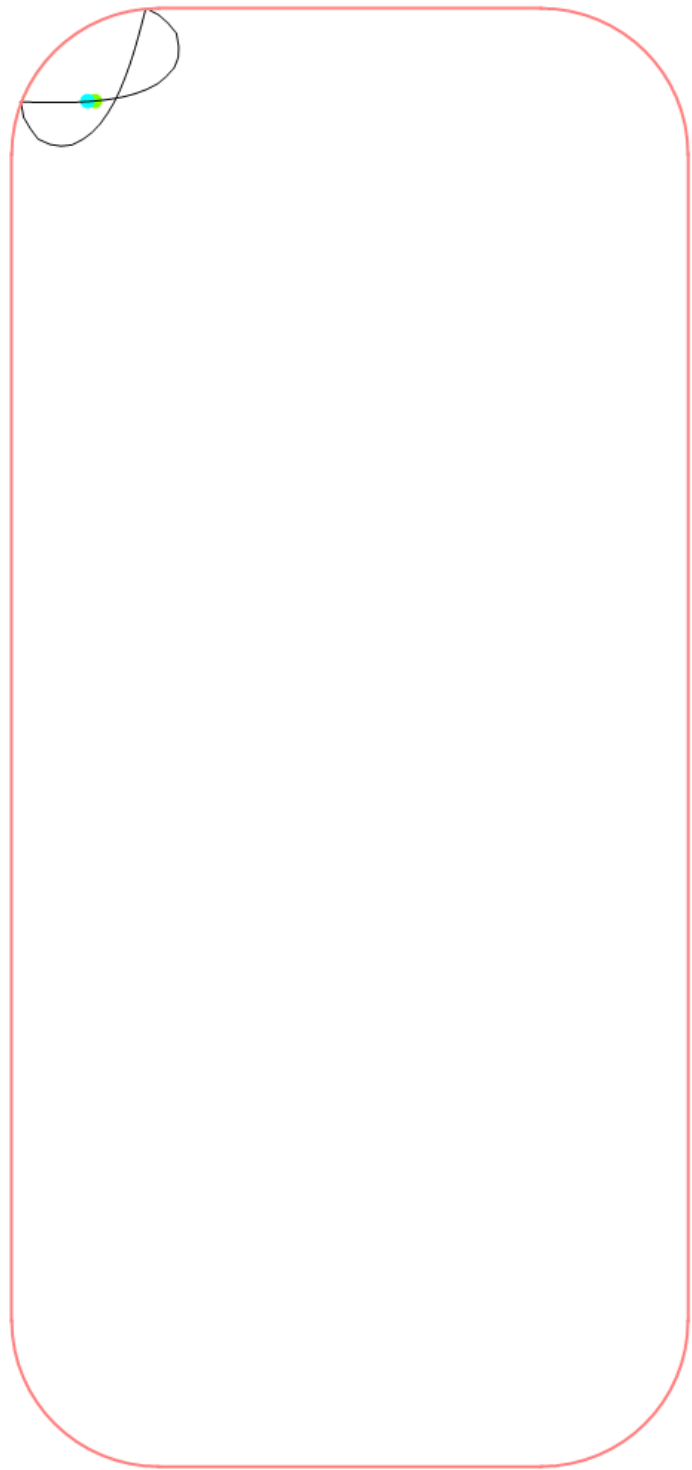

**Figure S3**

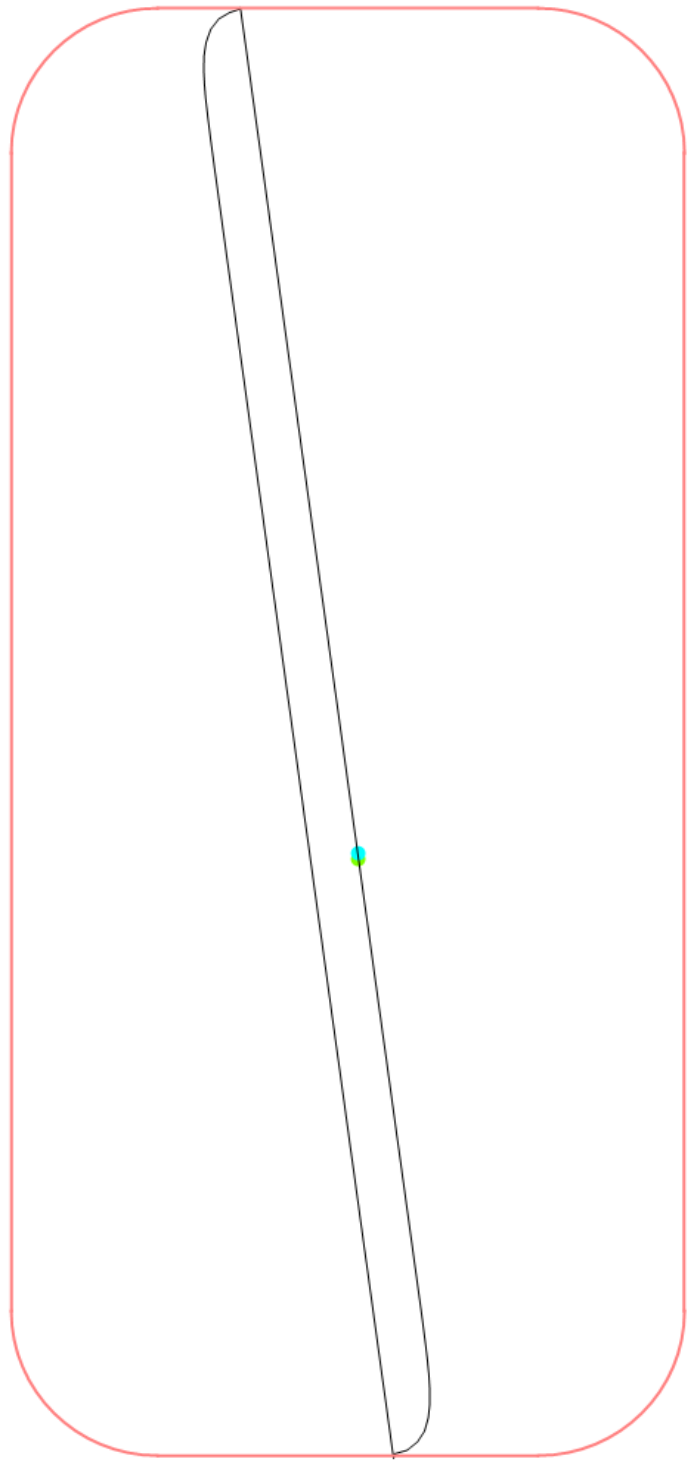